\documentclass[twocolumn,twoside,slac_two]{revtex4}
\usepackage{graphicx}
\usepackage{fancyhdr}
\usepackage{color}

\newcommand{\hessj}{HESS\,J1641$-$463}
\newcommand{\fourty}{HESS\,J1640$-$465}
\newcommand{\hess}{\textsc{H.E.S.S.}}
\newcommand{\g}{\ensuremath{\gamma}}%

\begin{document}

\title{Discovery of the VHE gamma-ray source \hessj}

%

\author{I. Oya, U. Schwanke, G. Spengler}
\affiliation{Institut f\"{u}r Physik, Humboldt-Universit\"{a}t zu Berlin, Newtonstrasse 15, D-12489 Berlin, Germany}
\author{M. Dalton}
\affiliation{Centre d' \'{E}tudes Nucl\'{e}aires de Bordeaux Gradignan, Universit\'{e} Bordeaux 1, CNRS/IN2P3 33175 Gradignan, France}
\author{B. Behera}
\affiliation{Deutsches Elektronen-Synchotron, DESY, Platanenallee 6, D-15738 Zeuthen,  Germany}
\author{P. Bordas}
\affiliation{Institut f\"{u}r Astronomie und Astrophysik, Universit\"{a}t T\"{u}bingen, Sand 1, D 72076 T\"{u}bingen, Germany}
\author{A. Djannati-Ata\"{i}}
\affiliation{APC, AstroParticule et Cosmologie, Universit\'{e} Paris Diderot, CNRS/IN2P3, CEA/Irfu, Observatoire de Paris, Sorbonne Paris Cit\'{e}, 10, rue Alice Domon et L\'{e}onie Duquet, 75205 Paris Cedex 13, France.}
\author{J. Hahn, V. Marandon}
\affiliation{Max-Planck-Institut f\"{u}r Kernphysik, P.O. Box 103980, D-69029 Heidelberg, Germany}
\author{For the H.E.S.S. Collaboration}

\begin{abstract}
A new TeV source, \hessj, has been serendipitously discovered
in the Galactic plane by the High Energy Stereoscopic System
(H.E.S.S.) at a significance level of 8.6 standard deviations. The
observations of \hessj\ were performed between 2004 and 2011
and the source has a moderate flux level of 1.7\% of the Crab Nebula
flux at E \textgreater 1 TeV. \hessj\ has a rather hard photon
index of 1.99 $\pm$ 0.13$\mathrm{_{stat}}$ $\pm$ 0.20$\mathrm{_{sys}}$. \hessj\ is positionally
coincident with the radio supernova remnant SNR G338.5+0.1, but no
clear X-ray counterpart has been found in archival Chandra
observations of the region. Different possible VHE production
scenarios will be discussed in this contribution.

\end{abstract}

\maketitle

\thispagestyle{fancy}


\section{H.E.S.S. analysis and results }
\begin{figure*}[t]
\centering
\includegraphics[width=110mm]{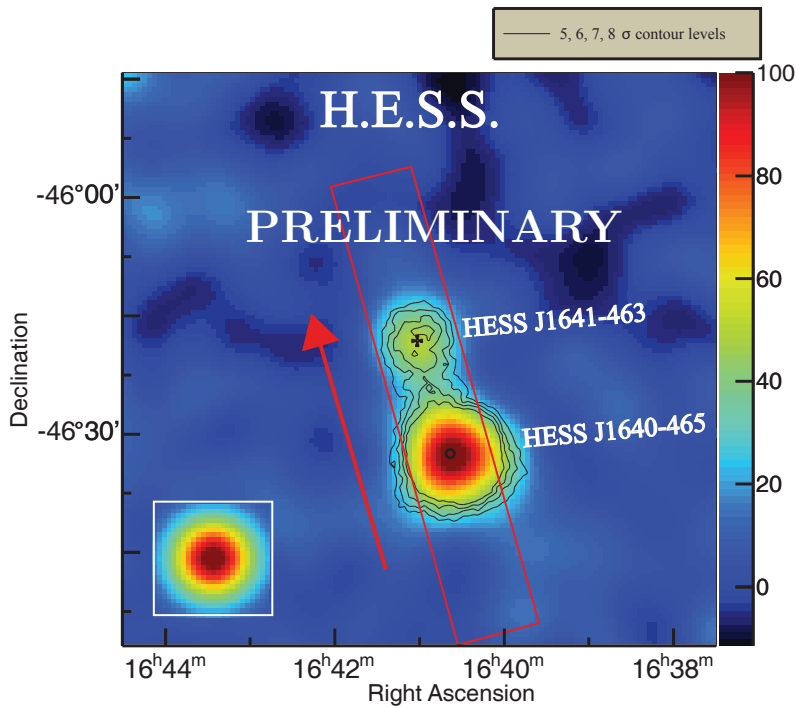}
\begin{picture}(0,0)
\put(-220,180){\color{white} \LARGE \bf PRELIMINARY}
\end{picture}
\caption{Smoothed map (smoothing radius 0.05$^{\circ}$) of excess events with
  energies E $>$ 4 TeV for the region around \hessj. The black
  solid contours indicate the significance of the emission at the
  5$\sigma$, 6$\sigma$, 7$\sigma$ and 8$\sigma$ levels. the box
  and arrow indicate the area and direction for the extraction of the
  profiles shown in Fig.~\ref{Fig2}. The inset illustrates the PSF of the
  instrument.} \label{Fig1}
\end{figure*}

H.E.S.S. is an array of five imaging Cherenkov telescopes situated in
the Khomas Highland in Namibia at an altitude of 1800 m above sea
level (see e.g. \cite{Bernloehr,Funk}). In the initial phase of the
H.E.S.S. project ({\it Phase I}), the array was composed of four 13 m
diameter Cherenkov telescopes, whereas in {\it Phase II} a single huge
dish with about 600 m$^{2}$ mirror area was added at the center of the
array, increasing the energy coverage, sensitivity and angular
resolution of the instrument \citep{HESSII}.  Due to the accumulation
of exposure with H.E.S.S. in the Galactic plane, complex VHE (Very
High Energy; E $>$ 100\,TeV) gamma-ray sources are better resolved and
frequently new structures appear. This is the case for a new source,
dubbed \hessj, that has been discovered near the strong gamma-ray
source HESS J1640-465 \citep{Aharonian2005}. Observations have been
performed between 2004 and 2011 with H.E.S.S. {\it Phase I} for a
total acceptance-corrected livetime of 72 h. The standard H.E.S.S. run
selection procedure has been used to select observations taken under
good weather conditions, and data have been analyzed with the Hillas
analysis technique \citep{Aharonian2006}, resulting in the detection
of a new source with a significance of 8.6$\sigma$ above 4 TeV (see
Fig.~\ref{Fig1}), \citep{LiMa}. The significance of the detection
corresponds to an excess of $68$ counts at the source best fit
position, the total number of ON- and OFF-source events being $\rm
N_{ON} = 107$ and $\rm N_{OFF} = 757$, respectively, with an on/off
sampling area factor of $\alpha = 0.05$.

\begin{figure*}[t]
\centering
\includegraphics[width=80mm]{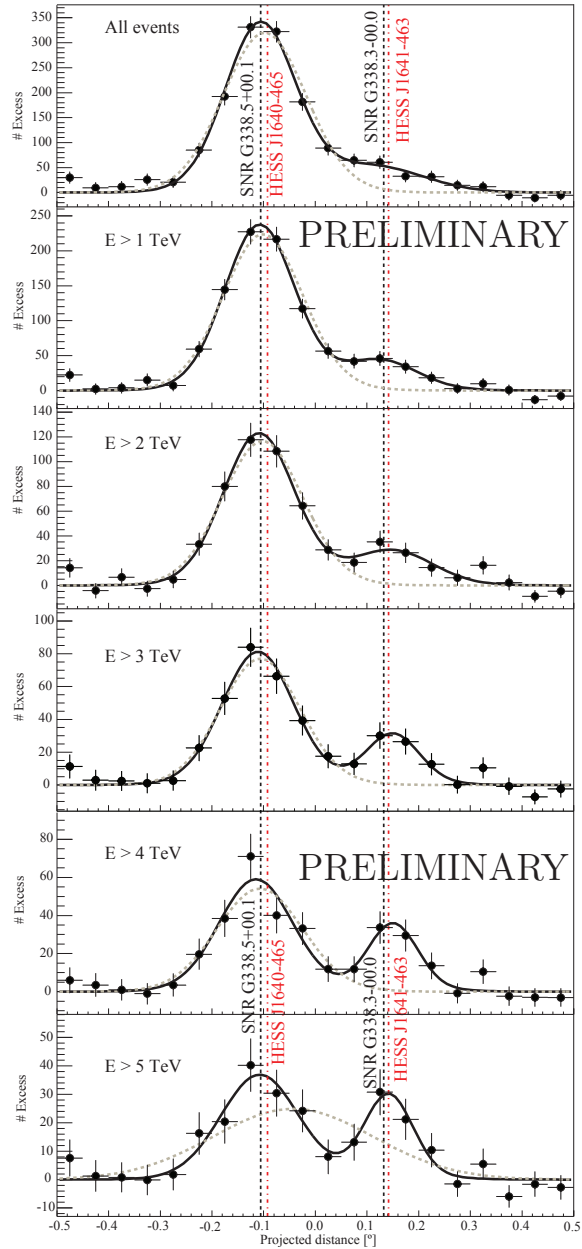}
\begin{picture}(0,0)
\put(-120,140){\Large PRELIMINARY}
\put(-120,380){\Large PRELIMINARY}
\end{picture}
\caption{Distribution of gamma-ray flux along the slice indicated in
  Fig.~\ref{Fig1}. Full lines model a double Gaussian convolved with
  the instrument PSF, (compared with a single Gaussian indicated with
  the dashed line). Vertical lines show the position of SNR 338.0+0.0,
  SNR G338.5+0.1, \hessj\ and \fourty.}
\label{Fig2}
\end{figure*}

The emission of \fourty/\hessj\ has been modeled with a double
Gaussian function convolved with the instrument point spread function
(PSF), showing a clear increase of the second source with increasing
energy (see Fig.~\ref{Fig2}). The preliminary best-fit position of
\hessj\ is found to be at RA: 16h 41m 1.7s $\pm$
3.1$\mathrm{s_{stat}}$ $\pm$ 1.9$\mathrm{s_{sys}}$, DEC =−46$^{\circ}$
18$'$ 11$''$ $\pm$ 35$''\mathrm{_{stat}}$ $\pm$ 20$''\mathrm{_{sys}}$
(J2000).

The time-averaged differential VHE $\gamma$-ray spectrum of the
source, has been derived using the forward-folding technique described
in \cite{Piron}. The spectrum is well fitted by a power-law function
with a photon index of 1.99 $\pm$ 0.13$\mathrm{_{stat}}$ $\pm$
0.20$\mathrm{_{sys}}$ for the energy range from 0.64 to 30\,TeV
($\chi$/d.o.f. equivalent value of 9.08/7 corresponding to a p-value
of 25\%). The flux density of \hessj\ is 1.9 $\pm$
0.2$\mathrm{_{stat}}$ $\times$ 10$^{-12}$ erg cm$^{-2}$ s$^{-1}$ above
1 TeV, corresponding to a 1.7\% of the Crab nebula flux at this
energy. The spectral index of \hessj\ is one of the hardests found so
far in a VHE \g-ray source.

No significant variability in the VHE $\gamma$-ray emission could be
established for \hessj. The integral flux is found to be constant
within errors over the \hess\ dataset. A fit of the period-by-period
light curve for energies above threshold with a constant value yields
a $\rm \chi^2/dof = 11.7 / 14$, with a p-value of $67\%$. No
variability can be seen either in other time binnings tested (from
year-by-year to run-by-run testing).

\section{Searching for a MWL counterpart}
\begin{figure*}[t]
\centering
\includegraphics[width=110mm]{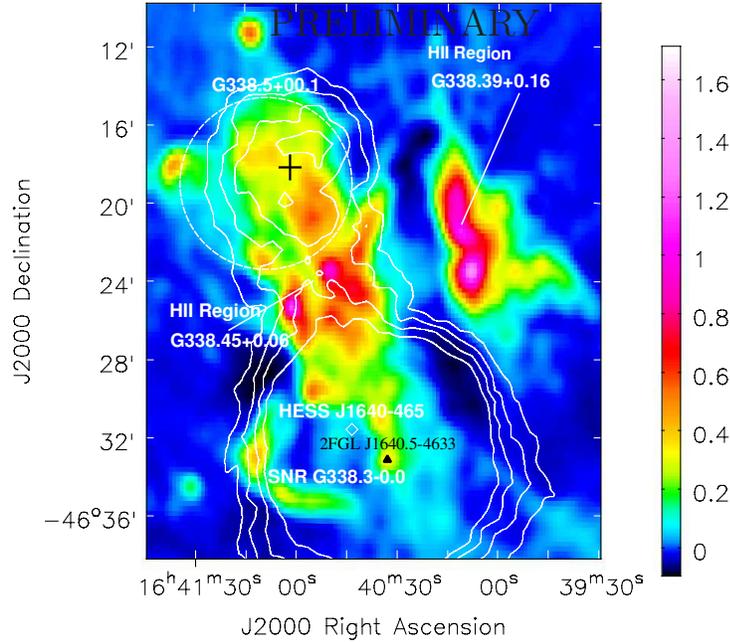}
\begin{picture}(0,0)
\put(-200,265){\Large PRELIMINARY}
\end{picture}
\caption{Radio view (MOST 843 MHz) of the region of interest.  The
  catalog position and extension of the SNR is indicated as a dashed
  white circle. The solid contours indicate the significance of the
  emission at E $>$ 4 TeV at the 5$\sigma$, 6$\sigma$, 7$\sigma$ and
  8$\sigma$ levels. The black cross indicates the best fit position of
  \hessj, and the black triangle the location of 2FGL
  J1640.5-4633.} \label{Fig3}
\end{figure*}

\hessj\ is found within the bounds of the Supernova Remnant
(SNR) G338.5+0.1 \citep{Green}, see Fig.~\ref{Fig3}. According to the
Green Catalogue of SNR \citep{Green}, SNR G338.5+0.1 has a roughly
circular morphology, and shows a flux density at 1 GHz of 12 Jy and an
angular size of 0.15$^{\circ}$. The distance to the SNR is estimated
to be $\sim$11 kpc \citep{Kothes}, indicating that it has a physical size of
about 30 pc. No X-ray source or massive star counterpart is found in
the different catalogues within 0.03$^{\circ}$ of the best fit
position of \hessj. No Fermi LAT source is located in the area,
besides the nearby 2FGL\,J1640.5-4633, associated with \fourty\
\citep{Fermi1640}.

\begin{figure*}[t]
\centering
\begin{minipage}{.5\textwidth}
  \centering
  \includegraphics[width=\linewidth]{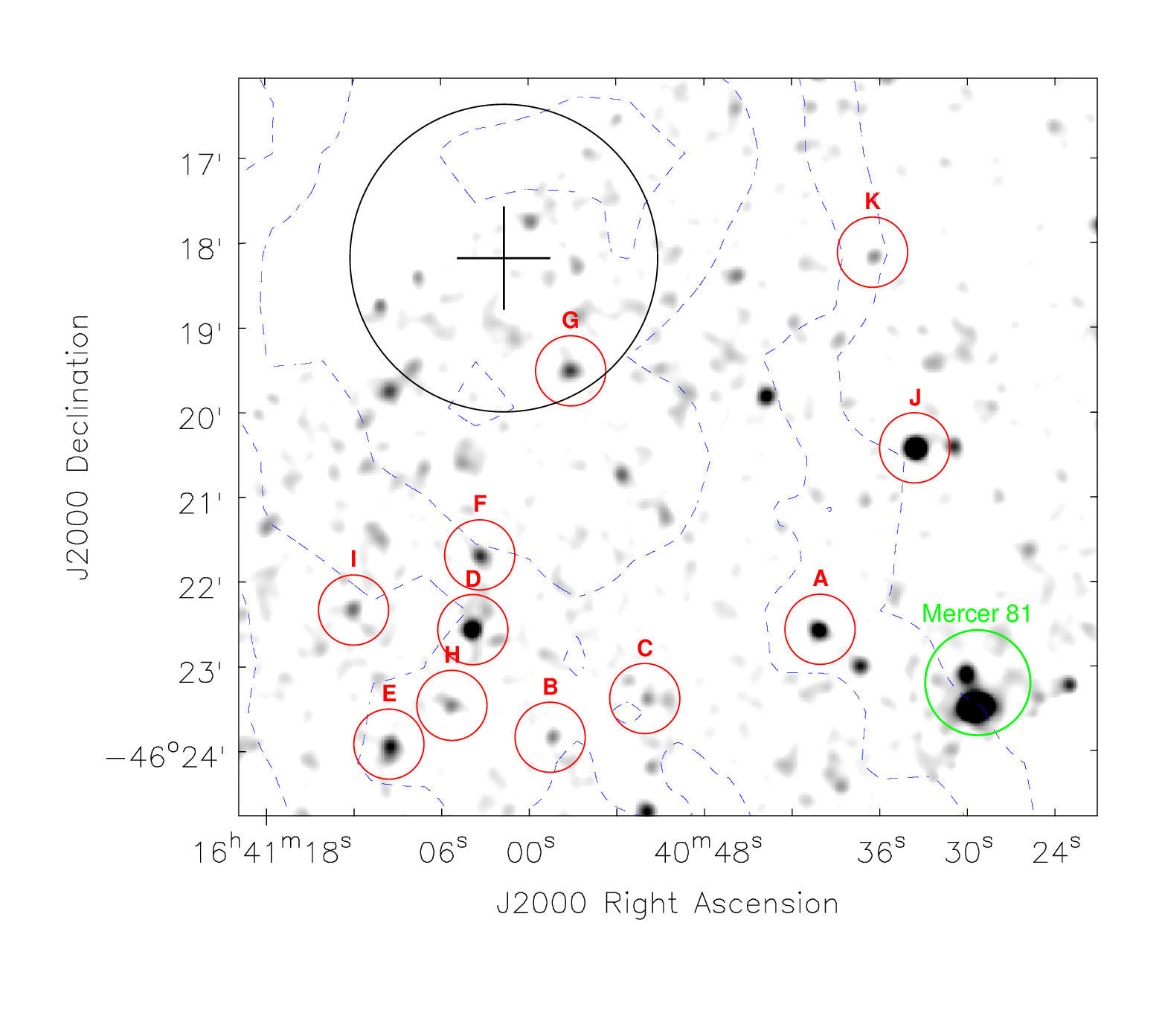}
  \end{minipage}%
\begin{minipage}{.5\textwidth}
  \centering
  \includegraphics[width=\linewidth]{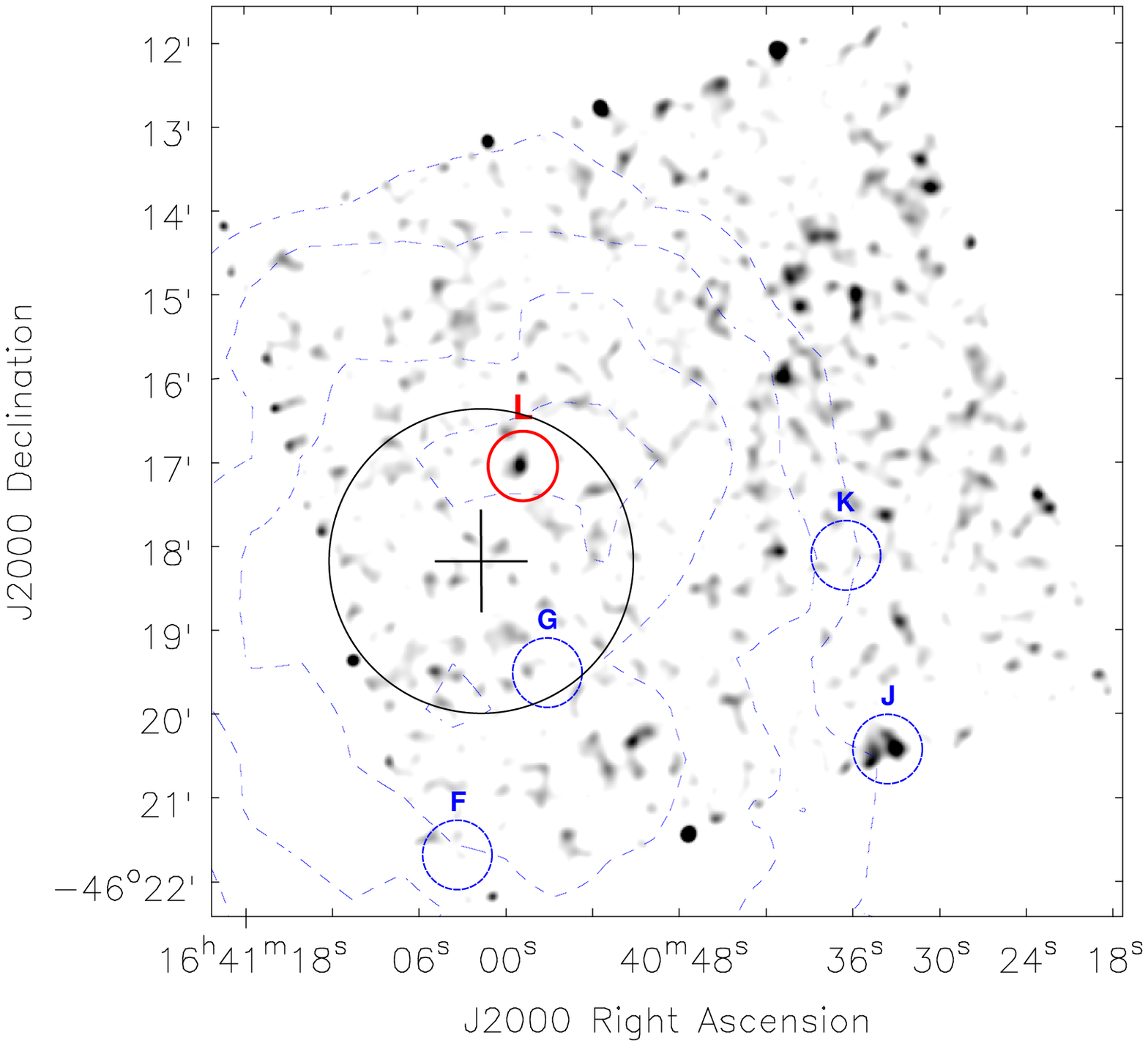}
  \begin{picture}(0,0)
    \put(-240,190){\Large PRELIMINARY}
    \put(-20,190){\Large PRELIMINARY}
  \end{picture}
\end{minipage}%
\caption{ Exposure corrected, smoothed (10 arcsec) image from Chandra
  Obs 11008 (left) and 12508 (right) of the area near \hessj\
  (the black cross indicating the best fit, the black circle a region of
  0.03$^{\circ}$ radius around it, and the dashed contours the
  significance of the emission at E \textgreater 4 TeV at the
  5$\sigma$ to 8$\sigma$ levels). The small circles indicate the
  detected faint X-ray sources.} \label{Fig4}
\end{figure*}

Two archival {\it Chandra} observations, both obtained with the
Advanced CCD Imaging Spectrometer (ACIS) detector, cover the region of
interest (ROI) (Tab \ref{Tab1}). These archival data have been
re-calibrated and analyzed with CIAO 4.4 and CALDB 4.4.7 versions. The
area at a distance smaller than 0.1$^{\circ}$ to the best fit position
of \hessj\ has been scanned with the CIAO \texttt{wavdetect},
\texttt{celldetect} and \texttt{vtpdetect} tools, revealing the
presence of 25 faint X-ray sources in the region. In order to reduce
the sample of revealed sources, a filter has been applied by
requesting a S/N \textgreater 3, and a positive hardness ratio of
counts with energies in the range of 2$-$10 keV over counts with
0.3$-$2 keV. After these selection criteria, the sample is reduced to
to 12 sources, and from these, two are located within 0.03$^{\circ}$
of the best-fit position of \hessj\ (see Fig.~\ref{Fig4} and tab
2.). The flux densities of these faint sources have been calculated
with the CIAO command \texttt{calc\_energy\_flux} for the 0.3$-$10 keV
energy range (see Tab. \ref{Tab2}).

In order to investigate a possible variability in the emission of the
found X-ray sources, a one-sample Kolmogorov-Smirnov (KS) test has
been used to calculate the probability, $P_{KS}$, for the null
hypothesis of a uniform flux for each found X-ray source (see
Fig.~\ref{Fig4}). For all the sources, $P_{KS}>$0.1 and therefore no
variability was found in the emission of any of them.

\section{Discussion}
The fact that \hessj\ is found within the bounds of a radio SNR
immediately suggests the possibility of the detection of a new VHE
SNR. However, the existing X-ray observations do not provide
additional support to this scenario due to the lack of detection of an
extended X-ray feature at the position of \hessj. In addition, the
larger extension of SNR G338.5+0.1 as compared to \hessj, and the
relatively old age of the SNR inferred from its physical size of
$\sim$30 pc suggests that the emission might not be necessarily
connected with the SNR but rather with a Pulsar Wind Nebula (PWN) at
its center, driven by an yet undetected pulsar. From energy
considerations, any of the faint X-ray sources of the area near
\hessj\ can be its counterpart in the PWN scenario. In particular, the
flux density level of the nearest found X-ray source, (source L) is a
factor $\sim$ 15 less energetic when compared with the energy flux
density of \hessj\ at the VHE energies. Yet, several associations of
``dark'' TeV sources and weak X-ray synchrotron PWN have been
established, were ratios of TeV/keV energy flux much larger have been
observed (e.g. HESS J1303-631 \citep{1303pwn}).

Another possibility could be that \hessj\ is a binary binary system,
similar to HESS J0632+057 \citep{Hinton}. Hard spectral indexes
($\Gamma \sim 1.2 - 2$) are found in those systems, similar to those
measured in \hessj. In a $\g$-ray binary scenario, however, a certain
degree of flux variability in both X-rays and TeV $\g$-rays is
expected, although this variability could appear in a relatively large
range of time-scales. In this regard, the lack of such variability
features does not permit to further constrain the $\g$-ray binary
scenario based on the temporal distribution of the TeV datasets. A
flux of $\sim$10$^{-14}$ erg cm$^{-2}$ s$^{-1}$ may be expected from
an X-ray faint binary system similar to HESS J0632+057 if located at
the estimated distance of SNR G338.5+0.1 (11 kpc). Massive stars that
could be tracers of the location of a \g-ray binary position were
investigated, but none were found. However, the lack of an obvious
optical counterpart could be explained due to the high optical
extinction expected given both the large distance and the position of
the source, close to the Galactic plane.

\section{Conclusions}
The excellent capabilities of H.E.S.S. have enabled the serendipitous
discover a new TeV source showing one of the hardest spectral indices
found so far. The inspection of astronomical catalogues and archival
X-ray observations do not provide an obvious counterpart to the VHE
source but reveal two potential, but weak, X-ray counterparts. The
spectral and morphological characteristics of \hessj, as well as its
location induce to think it could be most likely a SNR or a PWN, but
other source types like a binary cannot be excluded with the existing
data and therefore the source remains unidentified. Further
observations, in particular with high resolution X-ray instruments,
are encouraged for the proper identification of this new VHE source.

%


\begin{table}
\begin{center}
\caption{X-ray observations.}
\begin{tabular}{|l|c|c|c|c|}
\hline \textbf{Obs} & \textbf{Original} & \textbf{Exposure} & 
\textbf{Mode} & \textbf{Chips}\\
\textbf{ID} & \textbf{target} & \textbf{time} &  & \textbf{with ROI}\\
\hline 11008 & Mercer 81 & 40.0 ks & VFaint & I2 and I3 \\
\hline 12508 & Norma & 18.8 ks & VFaint & S3 \\
\hline
\end{tabular}
\label{Tab1}
\end{center}
\end{table}

\begin{table}[t]
\begin{center}
\caption{X-ray sources within the possible extension of \hessj.}
\begin{tabular}{|l|c|c|c|c|}
\hline \textbf{Source} & \textbf{RA} & \textbf{Dec} & 
\textbf{Net} & \textbf{Flux density}\\
\textbf{ID} & \textbf{(J2000)} & \textbf{(J2000)} & 
\textbf{counts} &  erg cm$^{-2}$ s$^{-1}$\\
\hline L & 16:40:58.9 & -46:17:02.8 & 15.2$\pm$5.1 & 1.5 10$^{-13}$ \\
\hline G & 16:40:57.2 & -46:19:30.7 & 20.7$\pm5.0$ & 4.2 10$^{-14}$ \\
\hline
\end{tabular}
\label{Tab2}
\end{center}
\end{table}


%



\bigskip 
\begin{acknowledgments}
The support of the Namibian authorities and of the University of
Namibia in facilitating the construction and operation of H.E.S.S. is
gratefully acknowledged, as is the support by the German Ministry for
Education and Research (BMBF), the Max Planck Society, the French
Ministry for Research, the CNRS-IN2P3 and the Astroparticle
Interdisciplinary Programme of the CNRS, the U.K. Science and
Technology Facilities Council (STFC), the IPNP of the Charles
University, the Polish Ministry of Science and Higher Education, the
South African Department of Science and Technology and National
Research Foundation, and by the University of Namibia. We appreciate
the excellent work of the technical support staff in Berlin, Durham,
Hamburg, Heidelberg, Palaiseau, Paris, Saclay, and in Namibia in the
construction and operation of the equipment.  

This research has made use of {\it Chandra} Archival data, as well as
the {\it Chandra} Source Catalog, provided by the {\it Chandra} X-ray
Center (CXC) as part of the {\it Chandra} Data Archive. This research
has made use of software provided by the {\it Chandra} X-ray Center
(CXC) in the application packages CIAO, ChIPS, and Sherpa. This
research has made use of the SIMBAD database, operated at CDS,
Strasbourg, France.
\end{acknowledgments}

\bigskip 

\end{document}